\documentclass[prl,aps,twocolumn,superscriptaddress,showpacs,footinbib]{revtex4-1}
\usepackage[colorlinks=true,linkcolor=black, citecolor=blue, urlcolor=blue, 
    unicode=true]{hyperref}

\usepackage{graphicx,amsmath,braket,tikz,amssymb,footmisc}

\begin{document}


\title{\texorpdfstring{Two-dimensional ferromagnetic spin-orbital excitations in the honeycomb VI$_{3}$}{}}

\author{H. Lane}
\affiliation{School of Physics and Astronomy, University of Edinburgh, Edinburgh EH9 3JZ, UK}
\affiliation{School of Chemistry and Centre for Science at Extreme Conditions, University of Edinburgh, Edinburgh EH9 3FJ, UK}
\affiliation{ISIS Pulsed Neutron and Muon Source, STFC Rutherford Appleton Laboratory, Harwell Campus, Didcot, Oxon, OX11 0QX, United Kingdom}
\author{E. Pachoud}
\affiliation{School of Chemistry and Centre for Science at Extreme Conditions, University of Edinburgh, Edinburgh EH9 3FJ, UK}
\author{J. A. Rodriguez-Rivera}
\affiliation{NIST Center for Neutron Research, National Institute of Standards and Technology, Gaithersburg, MD, USA}
\affiliation{Department of Materials Science and Engineering, University of Maryland, College Park, MD, USA}
\author{M. Songvilay}
\affiliation{School of Physics and Astronomy, University of Edinburgh, Edinburgh EH9 3JZ, UK}
\author{G. Xu}
\affiliation{NIST Center for Neutron Research, National Institute of Standards and Technology, Gaithersburg, MD, USA}
\author{P.M. Gehring}
\affiliation{NIST Center for Neutron Research, National Institute of Standards and Technology, Gaithersburg, MD, USA}
\author{J. P. Attfield}
\affiliation{School of Chemistry and Centre for Science at Extreme Conditions, University of Edinburgh, Edinburgh EH9 3FJ, UK}
\author{R. A. Ewings}
\affiliation{ISIS Pulsed Neutron and Muon Source, STFC Rutherford Appleton Laboratory, Harwell Campus, Didcot, Oxon, OX11 0QX, United Kingdom}
\author{C. Stock}
\affiliation{School of Physics and Astronomy, University of Edinburgh, Edinburgh EH9 3JZ, UK}

\date{\today}

\begin{abstract}

VI$_{3}$ is a ferromagnet with planar honeycomb sheets of bonded V$^{3+}$ ions held together by van der Waals forces.  We apply neutron spectroscopy to measure the two dimensional ($J/J_{c} \approx 17$) magnetic excitations in the ferromagnetic phase, finding two energetically gapped ($\Delta \approx k_{B} T_{c} \approx$ 55 K) and dispersive excitations.   We apply a multi-level spin wave formalism to describe the spectra in terms of two coexisting domains hosting differing V$^{3+}$ orbital ground states built from contrasting distorted octahedral environments. This analysis fits a common nearest neighbor in-plane exchange coupling ($J$=-8.6 $\pm$ 0.3 meV) between V$^{3+}$ sites.   The distorted local crystalline electric field combined with spin-orbit coupling provides the needed magnetic anisotropy for spatially long-ranged two-dimensional ferromagnetism in VI$_{3}$.

\end{abstract}


\maketitle

\textit{Introduction:} Order in two dimensions is forbidden by the Mermin-Wagner theorem~\cite{Mermin66:17,Hohenberg67:158,Mermin68:176,Nielsen80:22} in isotropic ferromagnets.  While Ising magnetic anisotropy has been theoretically shown to stabilize long-range magnetic order in two dimensions \cite{Khokhlachev,Bander}, achieving a strong enough single-ion anisotropy to overcome thermal fluctuations has been difficult to achieve in real materials.   The discovery of stable, spatially long-range ferromagnetism in two-dimensional materials~\cite{Burch18:563,Ajayan16:69,Duong17:11} such as CrI$_{3},$\cite{Chen,LiuPetrovic,McGuire,Wang,Kundu} Cr$_{2}$Ge$_{2}$Te$_{6}$,\cite{Gong,Carteaux} and Fe$_{3}$GeTe$_{2}$,\cite{Johansen,ChenYang,May,Calder,Zhang} has opened up the possibility of designing materials useful to spintronic applications~\cite{Ahn_2020,Dietl14:86} and for exotic two-dimensional physics to be explored such as topologically protected edge and surface modes \cite{Owerre,Pershoguba,Khaliullin13:111}.   We discuss two-dimensional ferromagnetism illustrating the effects of an orbital degree of freedom on the magnetic Hamiltonian and show that it can provide the necessary anisotropy to induce magnetic order.

\begin{figure}
  \includegraphics[width=85mm]{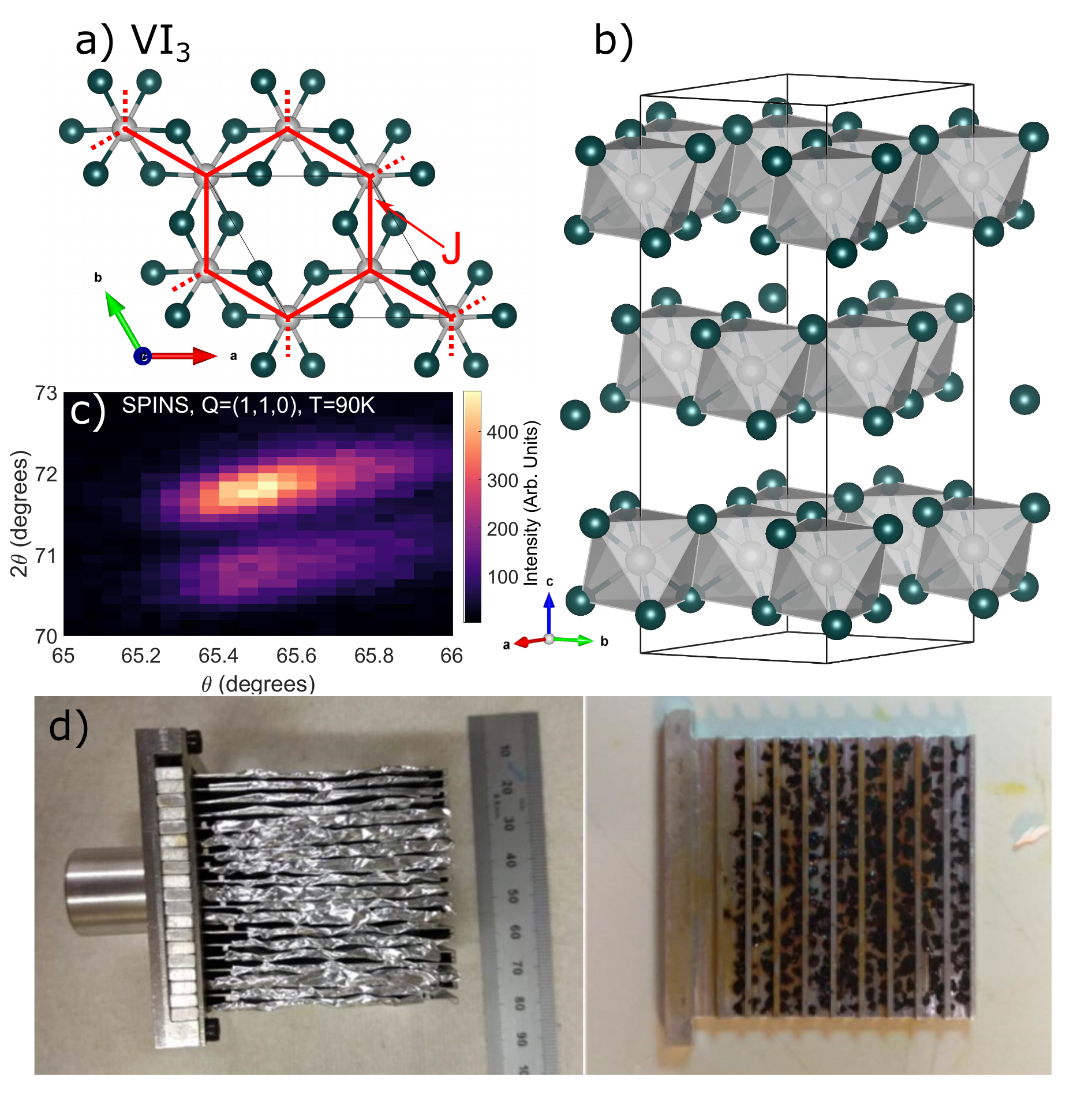}
 \caption{\label{Fig1} $(a)$ Structure of VI$_{3}$ in the $a$-$b$ plane showing the honeycomb lattice of V$^{3+}$ ions (gray) with an octahedral coordination of iodine ions (green). For this work, we take an $R\overline{3}$ unit cell.  $(b)$ VI$_{3}$ structure showing the stacking of two-dimensional sheets. $(c)$ (1,1,0) Bragg peak measured at SPINS, showing the existence of two domains at T=90 K. $(d)$ Aluminum sample mount showing co-aligned VI$_{3}$ crystals covered in Fomblin grease and mounted to a one of the nineteen panels. }
\end{figure}

VI$_{3}$ is unique amongst the two-dimensional van der Waals honeycomb ferromagnets as V$^{3+}$ ($S$=1) has degeneracy in the lower energy $t_{2g}$ orbitals \cite{Kong,Zhao21:103}, resulting in an entanglement of spin-orbital degrees of freedom that are coupled to the local structural environment~\cite{Gati,Dolezal}.  The structure of VI$_{3}$ (Fig. \ref{Fig1} $(a,b)$) is built upon V$^{3+}$ forming a layered honeycomb arrangement with an R$\overline{3}$ symmetry, stacked along $c$ with an $ABC$ arrangement.~\cite{Kong,Dolezal}  This stacking results in a rhombohedral superstructure~\cite{Marchandier21:xx}, though other symmetries have been discussed.~\cite{Tian,Son}  The $c$-axis stacking results in domains in large single crystals as evidenced by our scans of the (1, 1, 0) structural Bragg peak (indexed on an R$\overline{3}$ unit cell in Fig. \ref{Fig1} $(c)$) showing a splitting.  Given our interest in the two dimensional properties of V$^{3+}$, we consider an average R$\overline{3}$ structure here.  Below $T_{s}\approx 79\ \mathrm{K}$, a structural transition away from the R$\overline{3}$ is observed.~\cite{Tian,SI}

Magnetization and diffraction on VI$_{3}$ report a ferromagnetic transition($T_{c}\approx 50$ K)~\cite{Son,Kong,Tian,Dolezal,Gati,Liu,Lyu}, in agreement with Density Functional Theory~\cite{Subhan,An}. NMR\cite{Gati}, which probes the local V$^{3+}$ environment, has found the existence of two different ferromagnetic domains at low temperatures with differing local crystalline electric fields surrounding the V$^{3+}$ sites. This has further been supported theoretically~\cite{Yang_Khomskii,Huang} and also by diffraction~\cite{Dolezal}.  To understand the magnetic coupling and spin-orbital ground state, we apply neutron spectroscopy to probe the magnetic correlations at low temperatures.

\textit{Sample preparation:}  Over 1000 $\sim$ 1 mg single crystals of VI$_{3}$ were grown using chemical vapor transport~\cite{Juza} and edge-aligned using the hexagonal morphology (Fig. \ref{Fig1} $d$). The crystals were coated in hydrogen-free Fomblin oil on Al plates given their hydroscopic nature~\cite{Kratoch21:xx}.  

\begin{figure}
	\includegraphics[trim=1.7cm 3.3cm 1.2cm 2.0cm,width=80mm]{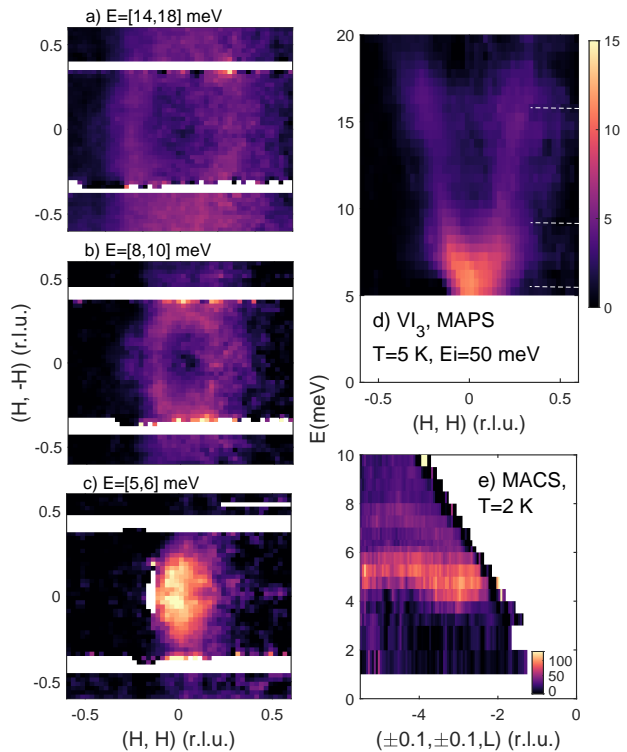}
	\caption{\label{ISIS}  $(a-c)$ T=5 K constant energy slices from MAPS. Energy integration ranges for each of the cuts in panels $(a-c)$ are given in square brackets. $(d)$ Momentum-energy slice illustrating dispersive modes from $Q$=0.  The locations of the constant energy slices are given by the dashed white lines. $(e)$ The excitations along the $c$-axis from MACS.  $L$ introduction is discussed in the SI.~\cite{SI}}
\end{figure}

\textit{Neutron Results:} Using the MAPS time-flight spectrometer (ISIS, Didcot, UK)~\cite{Ewings}, we first characterize the low temperature magnetic fluctuations in Fig. \ref{ISIS}. $E_{i}$ was set at 50 meV, with the Fermi chopper spinning at 200 Hz, giving an elastic energy resolution of 2.3 meV (FWHM).  The data were combined with the \texttt{Mantid/Horace} packages~\cite{Mantid,Horace,SI}. Figures \ref{ISIS} $(a-c)$ display constant energy cuts within the $a-b$ plane showing dispersive magnetic excitations.  Figure \ref{ISIS} $(d)$ shows a momentum-energy slice displaying the dispersive magnetic excitations up to the zone boundary at $\sim$ 20 meV.

\begin{figure}
  \includegraphics[trim=2.5cm 2.7cm 1.6cm 2.5cm,width=90mm]{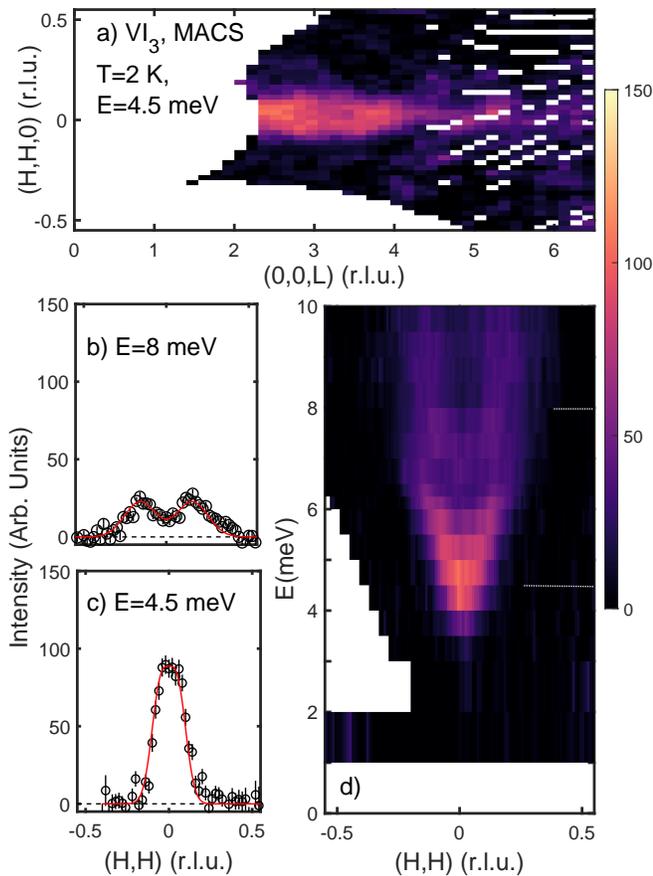}
 \caption{\label{NIST} $(a)$ Constant E=4.5 meV slice at 2 K from MACS, with background subtracted using methodology in Refs. \cite{Pasztorova,Stock18:121}.  $(b-c)$ constant energy cuts and $(d)$ momentum-energy slice integrating along $(0,0,L)$. The location of the constant energy scans are indicated by the dashed white lines. The $L$ introduction is discussed in the SI.~\cite{SI}}
\end{figure}

Low energy magnetic fluctuations were measured using the cold neutron spectrometer MACS (NIST, Gaithersburg, USA).\cite{Rodriguez}  The scattered neutron energy $E_{f}$ was fixed at 3.5 meV while the incident energy $E_{i}$ was varied, providing an elastic resolution of 0.25 meV (FWHM).  Fig. \ref{ISIS} $(e)$ displays the dispersion along $c$ illustrating little dispersion along this direction and affirming the two dimensional nature of the magnetic excitations and valdidating our consideration of a R$\overline{3}$ unit cell, neglecting the $ABC$ structural stacking.  This is confirmed in Fig. \ref{NIST} $(a)$ which plots a constant energy slice in the (HHL) plane illustrating a rod of scattering correlated in the (H,H,0) (in-plane) direction but extended along (0,0,L).  The decay of intensity with increasing momentum transfer along $(0,0,L)$ follows the V$^{3+}$ magnetic form factor~\cite{Brown_tables,SI}, implying the scattering is magnetic.  We note there is also a weak dispersion along $L$ (Fig. \ref{ISIS} $e$) which also results in a decay of intensity for a fixed energy transfer.  The magnetic in-plane coupling is illustrated in Figs. \ref{NIST} $(b)$ and $(c)$ with cuts along $(H,H)$ showing dispersive excitations at energies of 4.5 meV and 8 meV.  Fig. \ref{NIST} $(d)$ displays a $(0,0,L)$ integrated momentum-energy slice that shows two magnetic excitations dispersing along $(H,H)$ with gaps of $\sim$ 4 and $\sim$ 7 meV. 

Figure \ref{NIST} displays two gapped excitations indicative of local anisotropy which requires a finite energy to overcome.  However, the intensity variation with momentum transfer of the two modes is different.  The lower mode has a strong response near the zone center, but the intensity decays quickly away from $Q$=0 and is less dispersive.  The upper mode is fully mapped out in Fig. \ref{ISIS} and extends to higher energy and has a much more uniform intensity distribution across the Brillouin zone.  

The differing energy-momentum dependence of the two branches is suggestive of excitations from differing ground states.  Corroborating this is a comparison to the excitations in RbFe$^{2+}$Fe$^{3+}$O$_{6}$~\cite{Songvilay18:121} where the Fe$^{2+}$ (S=${2}$, $L$=2) and Fe$^{3+}$ (S=${5/2}$, $L$=0) display spatially long-range charge and orbital order.  In this case, two branches originating from the two different orbital iron ground states result in a weakly dispersive mode with intensity concentrated near the zone center and another mode that disperses more strongly throughout the zone with an even intensity distribution.  Motivated by this comparison and previous diffraction~\cite{Dolezal}, NMR~\cite{Gati}, and theoretical work~\cite{Yang_Khomskii} indicative of two orbital domains, we now investigate the magnetic excitations of VI$_{3}$ in the context of the spin-orbital properties of V$^{3+}$.  

\textit{Single-ion Hamiltonian:}  Given the near universality of the spatially localized crystalline electric parameters for transition metal ions, we first analyze the single ion V$^{3+}$ Hamiltonian with the goal of establishing the magnetic ground state of V$^{3+}$ that needs to be coupled in VI$_{3}$ and hence define the parameters to be extracted from experiment.  With the presence of an orbital degree of freedom and the low-temperature crystalline distortion and ferromagnetism, there are four single-ion Hamiltonian terms,

\begin{equation}
\mathcal{H}_{SI}=\mathcal{H}_{CEF}+\mathcal{H}_{SO}+\mathcal{H}_{dis}+\mathcal{H}_{MF}.
\end{equation}

\begin{figure*}
\includegraphics[width=175mm]{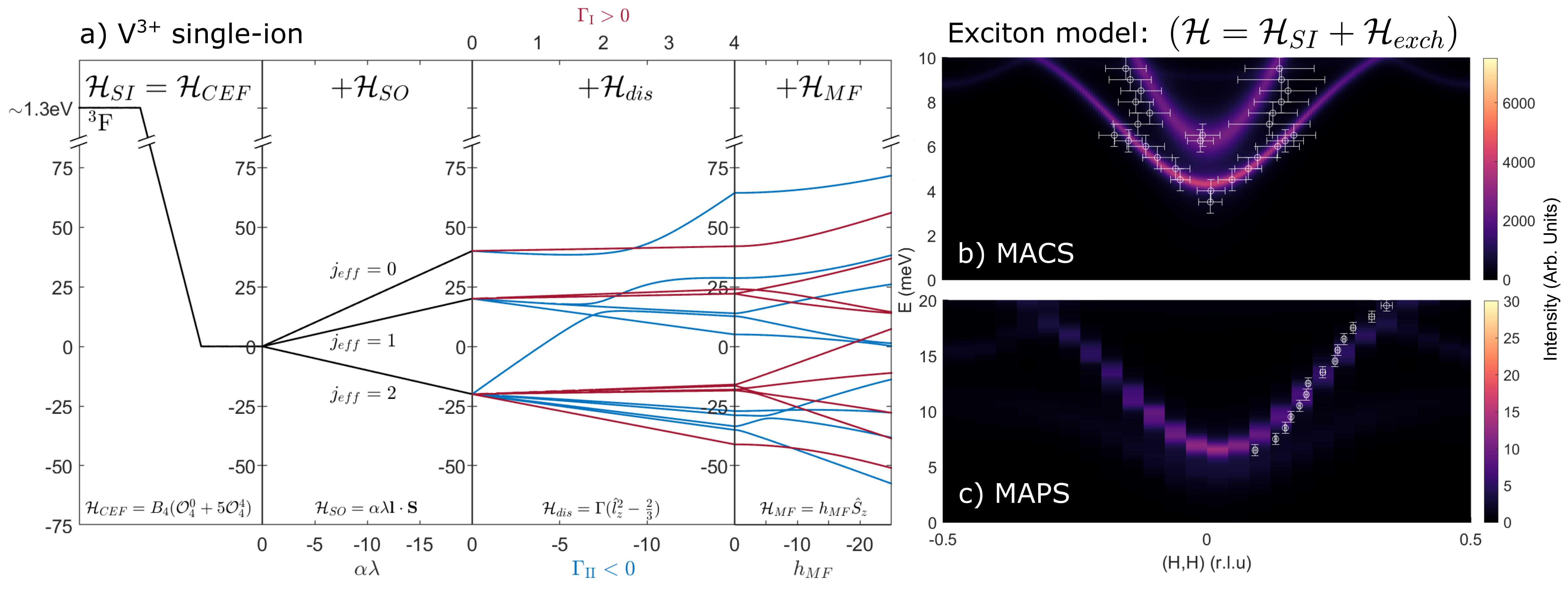}
\caption{\label{Fig4} $(a)$ Energy of V$^{3+}$ ion under a crystal field $\mathcal{H}_{CEF}$, spin-orbit coupling $\mathcal{H}_{SO}$, tetragonal distortion $\mathcal{H}_{dis}$, and mean molecular field $\mathcal{H}_{MF}$. Positive and negative distortions are shown in red and blue respectively. $(b)$ $S(\mathbf{Q},\omega)$ simulation of the MACS data (Fig. \ref{NIST} $(d)$) using the fitted values of exchange parameters. Overlaid data points were extracted from fitting Gaussian peaks to the data. $(c)$ Simulation of the MAPS data (Fig. \ref{ISIS} $(d)$) using Horace \cite{Ewings} to account for the finite integration ranges and detector coverage. Overlaid points were extracted from fitting Gaussian peaks to constant energy cuts.  }
\end{figure*}

\noindent This includes the octahedral crystalline electric field ($\mathcal{H}_{CEF}$), spin-orbit coupling ($\mathcal{H}_{SO}$), the structural distortion away from a perfect octahedron ($\mathcal{H}_{dis}$), and the local molecular field ($\mathcal{H}_{MF}$) imposed by ferromagnetic order.  We discuss each term in this Hamiltonian (Fig. \ref{Fig4} $a$) and its effect on the single-ion magnetic ground state.

\textit{$\mathcal{H}_{CEF}$-Octahedral field:}   In VI$_{3}$, the $d^{2}$ electrons forming a free ion $^{3}F$ are surrounded by six I$^{-}$ ions imposing a crystalline electric field on V$^{3+}$.  In terms of Stevens operators~\cite{Hutchings64:16,Bauer:book}, this lattice potential is written as $\mathcal{H}_{CEF}=B_{4}(\mathcal{O}_{4}^{0}+5\mathcal{O}_{4}^{4})$~\cite{Walter87:59} with the $^{3}F$ orbital ground state being energetically lowered by 360$B_{4}$ (Fig. \ref{Fig4} $a$), with an expected $B_{4}\sim$ 3.8 meV~\cite{McClure,AbragamBleaney}.  Refs. \cite{Yang_Khomskii,Huang} have alternatively discussed the single-ion properties of VI$_{3}$ using the strong crystal field approach~\cite{Stamokostas18:97,Khomskii}, whereby the crystalline electric field splits the five-fold $d$ orbital degeneracy into a ground state triplet $t_{2g}$, and excited doublet, $e_{g}$.  Either approach leads to a ground state projected ($L=\alpha l$) orbital triplet ($l$=1).  Given that other inorganic $3d$ metal complexes are typically in a high-spin state, we choose here the intermediate crystalline electric field basis with a projection factor $\alpha=-\frac{3}{2}$~\cite{moffitt59:2}.  The next excited state is 480$B_{4}\sim$1.8 eV~\cite{Tanabe54:9,Tanabe54:9_2,Cowley,KimNiO,Haverkort07:99,Larson07:99} which fixes the magnetic ground state of V$^{3+}$ to be $|l=1,S=1\rangle$. 

\textit{$\mathcal{H}_{SO}$-Spin-orbit coupling:}  The effect of spin-orbit coupling on the $|l=1,S=1\rangle$ ground state, with $\mathcal{H}=\alpha\lambda \mathbf{l} \cdot \mathbf{S}$, is shown in Fig. \ref{Fig4} $(a)$ and results in three levels with effective angular momentum values of $j_{eff}=0,1,2$.  For our analysis, we fix the spin-orbit coupling to the reported value of $\lambda$=12.9 meV \cite{AbragamBleaney}.  Given that V$^{3+}$ with $d^{2}$ electrons is less than half filled, it is expected that $\lambda>0$, implying $\alpha\lambda<0$.  The ground state is $j_{eff}$=2 separated from $j_{eff}$=1 by $2\alpha \lambda\sim$ 39 meV~\cite{AbragamBleaney,Griffith60:56}. 

\textit{$\mathcal{H}_{dis}$-Structural distortion:}  VI$_{3}$ is distorted from an ideal octahedron (Fig. \ref{Fig1} $b$).  Given orbitally driven transitions are primarily tetragonal~\cite{VanVleck,Goodenough,Koriki21:xx,Valenta20:xx},  we parameterize this as a distortion along $\hat{z}$ of the octahedra with $\mathcal{H}_{dis}=\Gamma_{\mathrm{I,II}}\left(\hat{l}_{z}^{2}-\frac{2}{3}\right)$ where $\Gamma$ is proportional to strain.   This additional energy term results in two possible orbital ground states, with $\Gamma_{\mathrm{II}} <0$ (flattened octahedra), an orbital ground state doublet while $\Gamma_{\mathrm{I}}>0$ (elongation) is a ground state singlet.  These two scenarios are shown in Fig. \ref{Fig4} $(a)$ in different colors. In the strong crystal field basis~\cite{Yang_Khomskii} one ground state is defined as a $d_{xz},d_{yz}$ doublet and a second with the $d_{xy}$ ground state singlet with one of the higher energy $d_{xz},d_{yz}$ orbitals occupied.  Given results in Refs.\cite{Buyers71:4,Cowley73:6,SarteCoO,SarteCRO}, we expect $|\Gamma|\sim$ 10 meV.   

\textit{$\mathcal{H}_{MF}$-Molecular Field:}  The final $\mathcal{H}_{SI}$ term is the molecular field present in the $T<T_{c}\sim 50$ K ferromagnetic phase from neighboring ordered spins inducing a Zeeman field on a V$^{3+}$ site.  The $\mathcal{H}_{MF}=h_{MF}\hat{S}_{z}$ term splits the degenerate spin-orbit levels and is fixed by the spin exchange which induces a molecular field $h_{MF}=\sum_{j}\mathcal{J}_{ij}\langle \hat{S}_{j}^{z}\rangle=3JS$ (Fig.\ref{Fig1} $a$).  Ferromagnetic exchange is expected based on 90$^{\circ}$ bonds between nearest V$^{3+}$ neighbors and validated by calculations~\cite{Yang_Khomskii}.  Molecular orbital calculations~\cite{Zhao21:103} predict $J\sim$ -7 meV, implying $h_{MF}\sim$ -20 meV.  This is of a similar magnitude to the spin-orbit coupling and induces many single-ion levels with a similar energy scale (Fig. \ref{Fig4}).     

\textit{Multi-level spin-waves:} The dispersive excitations shown in Figs. \ref{ISIS} and \ref{NIST} are indicative of coupled V$^{3+}$ ions with the Hamiltonian $\mathcal{H}=\mathcal{H}_{SI}+\mathcal{H}_{exchange}$, where $\mathcal{H}_{exchange}=\sum_{j}\mathcal{J}_{ij}\hat{S}_{i} \cdot \hat{S}_{j}$, describes an isotropic Heisenberg interaction between neighboring V$^{3+}$ ions.   The usual method of parameterizing such excitations is based on standard spin-wave theory where transverse deviations of an angular momentum vector of fixed magnitude are considered.  This is based on a ground state, energetically separated from other single-ion levels and is a valid approximation in many compounds with an orbital degeneracy~\cite{Kim,SalaBaIr,Edwards,Sarte18:98,Wallington15:92} where spin-orbit coupling is a perturbation and is parameterized through anisotropic terms~\cite{Yosida}.   With the presence of spin-orbit coupling of a similar magnitude to the exchange coupling, as in VI$_{3}$, this approach is not valid due to the mixing (Fig. \ref{Fig4} $a$) of single-ion spin-orbit levels~\cite{Feldmaier20:2} and necessitates a multi-level approach to the excitations.  Below, we apply such a methodology based on single-ion eigenstates where anisotropy terms are incorporated explicitly through the single-ion Hamiltonian described above.  
    
We fit Figs. \ref{ISIS} and \ref{NIST} with three parameters - $J$ and $\Gamma_{I,II}$ with other single-ion terms fixed to the literature values as described above (note $H_{MF}$ is fixed $J$). We use the Green's function equation of motion~\cite{Buyers,SarteCoO,SarteCRO} in terms of the eigenstates of $\mathcal{H}_{SI}$ to calculate the neutron response via the fluctuation-dissipation theorem $S(\mathbf{Q},\omega)\propto-f(\mathbf{Q})^{2}\mathrm{Im}\left(G(\mathbf{Q},\omega)\right)$~\cite{SI}.   This is consistent with other multi-level spin-wave theories.~\cite{Muniz14,Hasegawa12:81}  Within the random phase approximation, the transverse Green's functions for nearest neighbor coupling is,

\begin{align}
	\label{RPA}
	G_{\mu \nu}^{+-}(\mathbf{Q},\omega)=&g_{\mu}^{+-}(\omega)+g_{\mu}^{+-}(\omega)\mathcal{J}_{\mu\nu}(\mathbf{Q})G_{\mu\nu}^{+-}(\mathbf{Q},\omega)
\end{align}

\noindent where $\mathcal{J}_{\mu \nu}(\mathbf{Q})=\sum_{ij}J_{\mu\nu}e^{i\mathbf{Q}\cdot\mathbf{\delta}_{ij}}$ is the Fourier transform of the exchange interaction between nearest sites $\nu$ and $\mu$, and $g^{\alpha\beta}_{\mu}$ is the single-site susceptibility, defined as

\begin{equation}
g^{\alpha\beta}_{\mu}(\omega)=\sum\limits_{mn}\frac{\bra{m}\hat{S}^{\alpha}_{\mu}\ket{n}\bra{n}\hat{S}^{\beta}_{\mu}\ket{m}}{\omega-\left(\omega_{n}-\omega_{m}\right)}.
\end{equation}

\noindent The energies, $\omega_{n}$, are the eigenvalues of $\mathcal{H}_{SI}$, with $\ket{n}$ the single ion eigenstates.   VI$_{3}$ exhibits ABC stacking along $c$ (Fig. \ref{Fig1} $b$) \cite{Tian} requiring six sites $\mu,\nu=\{ 1,2,...,6\}$.

Based on Refs. \cite{Yang_Khomskii,Gati,Huang,Dolezal}, we consider two domains with oppositely distorted octahedra - $\Gamma_{\mathrm{I}}>0$ and $\Gamma_{\mathrm{II}}<0$.  For simplicity we fix the volume ratio $\Gamma_{I}/\Gamma_{II}$=1. Fig. \ref{NIST} $(a)$ and Fig. \ref{ISIS} $(e)$ indicate $J/J_{c}\approx 17$, therefore we neglect coupling along $c$, considering the nearest-neighbor in-plane exchange $J$ equal in both domains.  In terms of the momentum-energy structure of the magnetic excitations, the parameter $J$ tunes the dispersion of the magnetic modes and $\Gamma_{I,II}$ controls the size of the gap of the two excitations in Fig. \ref{NIST}.  Including more complex structural deviations has the effect of changing this gap size.~\cite{SI}  Akin to anisotropy terms incorporated into conventional spin-wave theory, $\Gamma_{I,II}$ describe the effects of the local single ion anisotropy from a distortion away from a perfect octahedral environment. 

Fig. \ref{Fig4} displays a three parameter fit with $J=-8.6$ $(\pm0.3)$ meV, $\Gamma_{\mathrm{II}}=-13.7$ $(\pm0.5)$ meV and $\Gamma_{\mathrm{I}}=3.4$ $(\pm0.02)$ meV. The upper mode is from the domain with a flattened octahedron (domain II) and the lower from elongation (domain I).  Despite the different energy bandwidths of the two modes, a common value of the nearest-neighbor $J$ is sufficient to describe the dispersion in both domains with the different dispersion bandwidths originating from the contrasting orbital ground states. The multi level spin-wave model captures the rapid intensity decay of the lower mode away from the zone center, however, we do not observe any intensity near the zone boundary in experiment,  in disagreement with model calculations.  This can be understood by finite lifetime effects due to disorder which has been both theoretically and experimentally found to disproportionately affect shorter wavelength excitations away from the magnetic zone center.~\cite{Helton,Halperin,Zhitomirsky,Chou}  This indicates stronger disorder for orbitally singlet V$^{3+}$ (domain I-elongation).  The stability of a flattening (domain II) of the octahedron around the V$^{3+}$ site is consistent with results found for other V$^{3+}$ compounds.~\cite{Tchernyshyov,Onoda03:15,Reehuis03:35}  Two distinct V$^{3+}$ domains, with one disordered, is also consistent with NMR results.~\cite{Gati} 

The multi-level model coupling single-ion states determined by spin-orbit coupling, distorted octahedra, and a molecular field results in gapped excitations consistent with the data with three parameters - $\Gamma_{I,II}$ and one exchange constant $J$.   This is in contrast with traditional spin-wave theory that would require two very different exchange parameters, for the differing domains, with the ratio scaling with the magnon bandwidths.  Such a large difference in exchange constants is difficult to justify through the local bonding environments and small deviations away from an average $R\overline{3}$ unit cell.  

The energy cost of excitations is determined by the energy gap at $Q$=0.  This is $\approx$5 meV=58 K, similar to the Curie temperature in VI$_{3}$, which defines ferromagnetic order.  This anisotropic gap, which facilitates magnetic order, originates from spin-orbit coupling.  We note that other two dimensional van der Waals magnets which lack spin-orbit coupling do not display spatially long-range order with NiGa$_{2}$S$_{4}$ an example.~\cite{Nakatsuji05:309, Stock10:105,Nakatsuji10:79,Nambu15:115}  The situation is different in  CrI$_{3}$~\cite{Chen} and CrBr$_{3}$~\cite{Samuelsen71:3} where Cr$^{3+}$ lacks an orbital degeneracy.   It is interesting that CrI$_{3}$ has a large Curie temperature, but is comparatively three-dimensional in terms of the magnetic exchange coupling~\cite{Chen} and critical properties.~\cite{LiuPetrovic,Liu19:9,Liu}  Spin-orbit coupling therefore can provide a route for creating a strong enough anisotropy that magnetic order is stable in two dimensions. 

In summary we have presented a neutron spectroscopy study of the effects of an orbital degree of freedom on the honeycomb van der Waals ferromagnet VI$_{3}$. We have parameterized these two modes in terms of two oppositely distorted domains and have presented multi spin-orbit level calculations to model the inelastic neutron scattering response with good agreement.  

\begin{acknowledgements}

The authors thank W.J.L. Buyers and P.M. Sarte for discussions and acknowledge funding from the EPSRC and STFC. Access to MACS was provided by the Center for High Resolution Neutron Scattering, a partnership between the National Institute of Standards and Technology and the National Science Foundation under Agreement No. DMR-1508249.  Experiments at the ISIS Pulsed Neutron and Muon Source were supported by beamtime allocation RB2010594 from the Science and Technology Facilities Council. H. L. was co-funded by the ISIS facility development studentship programme. 

\end{acknowledgements}


%

\end{document}


\title{Supplemental Material}
\maketitle

\section{Experimental Setups}

Three different neutron experiments were performed to support the data presented in the main text and to characterize the bulk samples and to also study the magnetic dynamics.  \\

$\textit{SPINS:}$  Initial structural studies were performed on a single crystal sealed in an aluminum can under helium gas, aligned such that Bragg reflections of the form (HHL) lay within the horizontal scattering plane.  The sample was cooled in a closed cycle displex refrigerator.  On SPINS, a vertically focussed PG(002) monochromator was used to select an incident energy of E$_{i}$=5.0 meV.  A flat PG(002) analyzer was used to fix the final energy to the elastic scattering condition with E$_{f}$=5.0 meV.  Cooled Beryllium filters were placed both in the incident and scattering beams to reduce higher order contamination.  The beam collimation sequence was set to $guide$-80-$Sample$-80-$open$. \\

\begin{figure}[h]
\includegraphics[trim=5cm 7cm 5cm 7cm,width=0.33\textwidth]{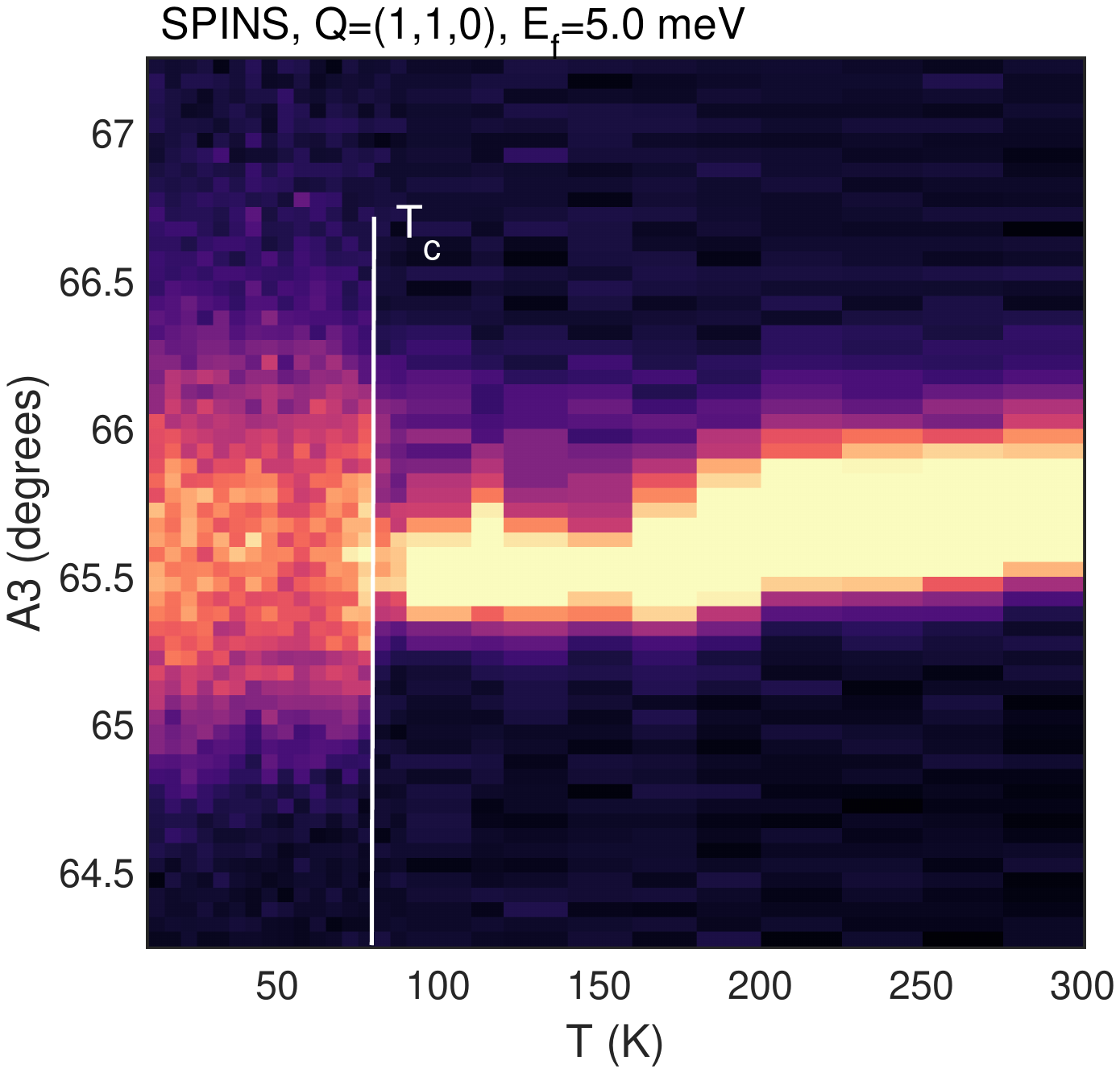}
\caption{\label{Bragg_temp} Temperature dependence of the (1,1,0) Bragg peak, as scanned in the sample angle denoted as $A3$, showing a transition at $T_{c}\approx 80$K.}
\end{figure}

$\textit{MACS:}$  Spectroscopy measurements were initially performed on MACS (NIST) \cite{Rodriguez} with the sample aligned such that reflections of the form (HHL) lay within the horizontal scattering plane.  The sample was cooled in an Orange cryostat to temperatures as low as 2 K.   A double focussed PG(002) monochromator was used to select an incident energy and focus the beam onto the sample.  20 PG(002) double bounce analyzers were then used to select a final energy of E$_{f}$=3.5 meV.  Cooled Beryllium filters were used on the scattered side of the sample to remove higher order contamination from the sample. \\

$\textit{MAPS:}$  To study the dynamics at higher energies beyond the energy range available at MACS, MAPS at the ISIS spallation neutron source was used \cite{Ewings,MAPSdata}.  An incident energy of E$_{i}$=50 meV was used with a Fermi chopper spinning at 200 Hz.  The sample was aligned such that Bragg reflections of the form (HHL) lay within the horizontal plane.  The sample was cooled in a top loaded closed cycle refrigerator to 5K.  To probe the ferromagnetic position and excitations throughout the Brillouin zone, the sample was rotated and then the data combined using the Horace data analysis package.

\section{Characterization}

To confirm the structural transition and characterize the structural domains in our samples we studied the (1,1,0) Bragg peak using SPINS at NIST as discussed above.  The (1,1,0) Bragg peak was measured as a function of temperature on the SPINS spectrometer (NIST, MD).   The results of the temperature dependence are shown in Fig. \ref{Bragg_temp} of one of the structural domains illustrating a transition at $\sim$ 80 K consistent with reports in the literature.

\section{Crystal synthesis}

Single crystals of VI$_{3}$ were synthesized using the chemical vapor transport technique \cite{Juza}.  Sealed quartz ampoules with outer diameter 18 mm and inner diameter of 16 mm were loaded with 3 g of Vanadium and Iodine in stoichiometric quantities.  Approximately 5\% excess of iodine, by mass, was included to act as a transport agent.  The vanadium powder was initially pumped to less than 10$^{-5}$ Torr using a turbo pump to ensure dryness before the iodine was loaded.  The combined reagents were then chilled and pumped to 5 $\times$ 10$^{-3}$ Torr using an oil based mechanical pump to avoid damage to the blades of the turbo pump.  The tubes were sealed to be a length of $\sim$ 15 cm and put into a 3-zone furnace such that one end was at 400 $^{\circ}$C and the other end was at 350$^{\circ}$C.  A chiller was used to further cool one end of the furnace to increase the temperature gradient.  The temperature gradient was initially inverted for 12 hours to clean one end of the ampoules.  12 ampoules were loaded for each crystal growth run which lasted 10 days.  The ampoules were then removed at high temperature with one end cooled using compressed air on removal from the three-zone furnace.   A variety of crystal sizes resulted in size up to a maximum of 5 mm $\times$ 5 mm, with a depth of less than 1mm. \\

\begin{figure}[h]
\includegraphics[trim=0cm 0cm 0cm 0cm,width=0.67\textwidth]{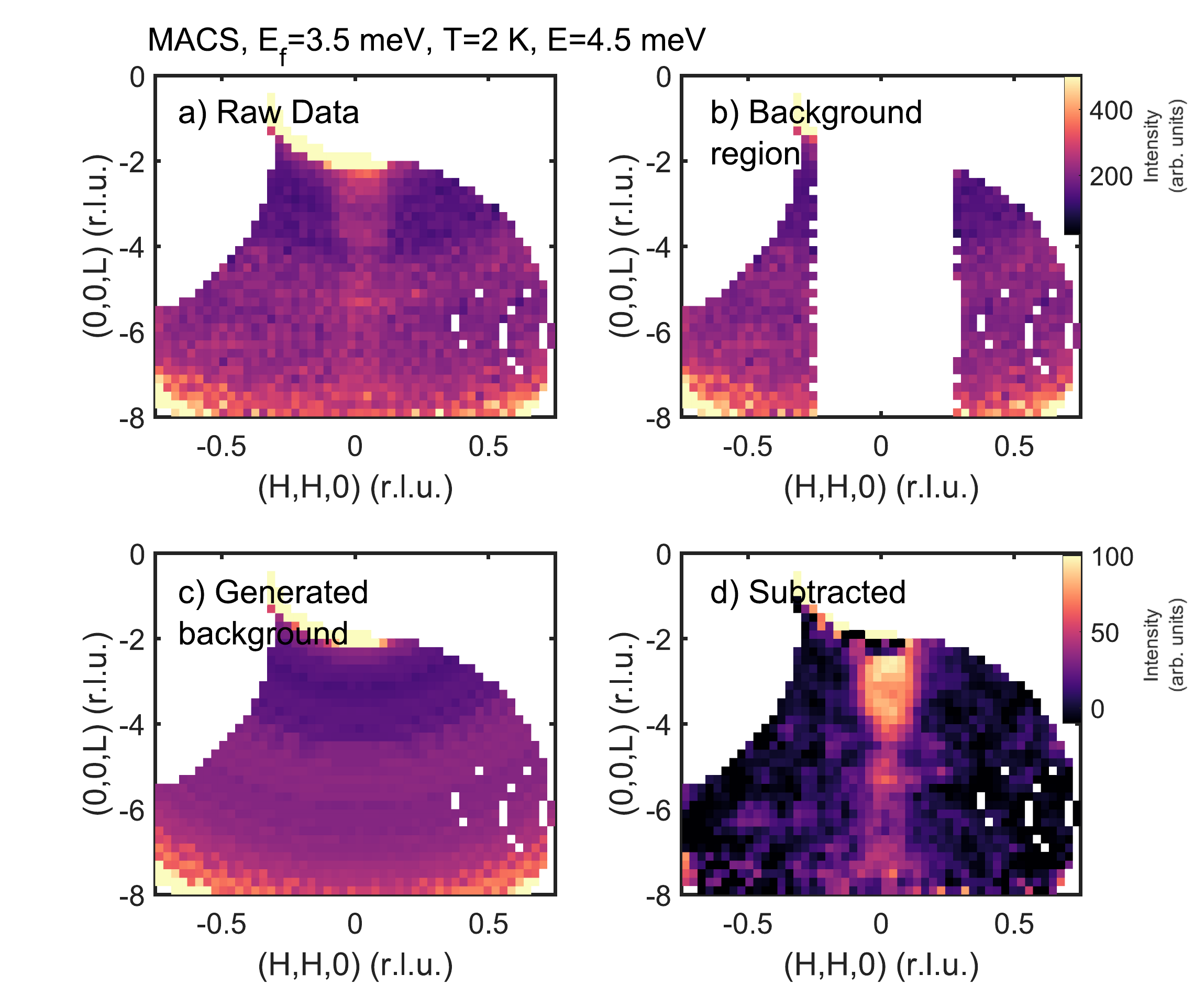}
\caption{\label{MACS_back} The background subtraction methodology applied to the MACS neutron inelastic scattering data. $(a)$ illustrates the raw data and $(b)$ the selected background region.  $(c)$ shows the resulting averaged generate background used to subtract from the data shown in $(d)$.}
\end{figure}

The samples were found to degrade under air on removal from the evacuated quartz ampoule.  To mitigate this the crystals were immediately soaked in Fomblin grease on removal and then attached to aluminum plates with shelves using Fomblin.  The samples were then kept in a vacuum oven purged with argon and filled with desiccant.  A vacuum was not used as it was found to pump off the Fomblin and degrade the samples.  To further preserve the samples, the shelved aluminum plates with the VI$_{3}$ crystals were wrapped with aluminum foil (also coated with Fomblin grease). \\

Two different sample mounts were created for the MACS (NIST) and MAPS (ISIS) experiments respectively.  The MACS experiment involved approximately 12 ampoules to construct a sample mount with plates $\sim$ 25 $\times$ 25 mm in size.  The goal of the higher energy MAPS experiment was to track the higher energy mode, which was found at MACS to be considerably weaker in intensity.  A larger sample mount, shown in the main paper, was constructed based on more than 60 ampoules with a larger surface area of 50 $\times$ 50 mm to take advantage of the larger neutron beam on MAPS.  We note that one large ($\sim$ 5 mm in length) single crystal was used for characterization using the SPINS cold  triple-axis spectrometer at NIST outlined above.

\begin{figure}[h]
\includegraphics[trim=0cm 1cm 0cm 0cm,width=0.6\textwidth]{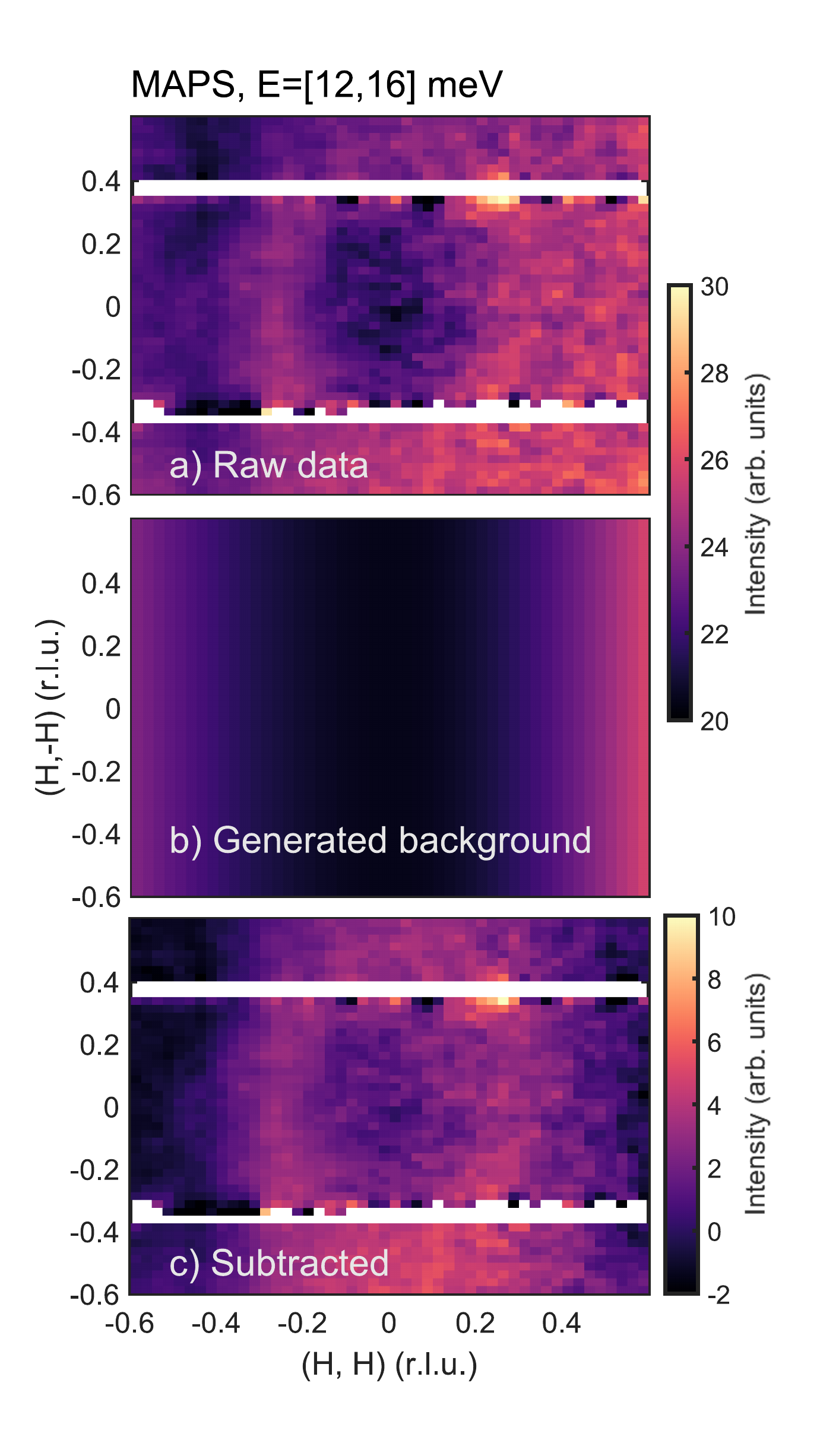}
\caption{\label{Back_MAPS} The background subtraction methodology applied to the MAPS dataset.  $(a)$ shows the raw uncorrected data and $(b)$ shows the estimated background from the fitting procedure discussed in the text.  $(c)$ shows the subtracted data.}
\end{figure}

\section{Background subtraction}

Given the plate-like sample mount and complex sample geometry on MACS and MAPS, we needed to subtract off the background applying different methodologies.  These are outlined here. The sample mounts for spectroscopic measurements consisted of aluminum plates with the samples attached with hydrogen-free Fomblin grease.  Fomblin consists of polymers of $-CF_{2}-$ and does not have hydrogen and therefore reduces the background from incoherent scattering that would be present with other adhesives. However, we note that the Fomblin grease does have an incoherent cross section and this prevented us from determining a reliable estimate of the sample mass from the incoherent elastic line, despite the large incoherent cross section of vanadium.\\

\begin{figure}[h]
	\includegraphics[trim=5cm 7cm 5cm 7cm,width=0.5\textwidth]{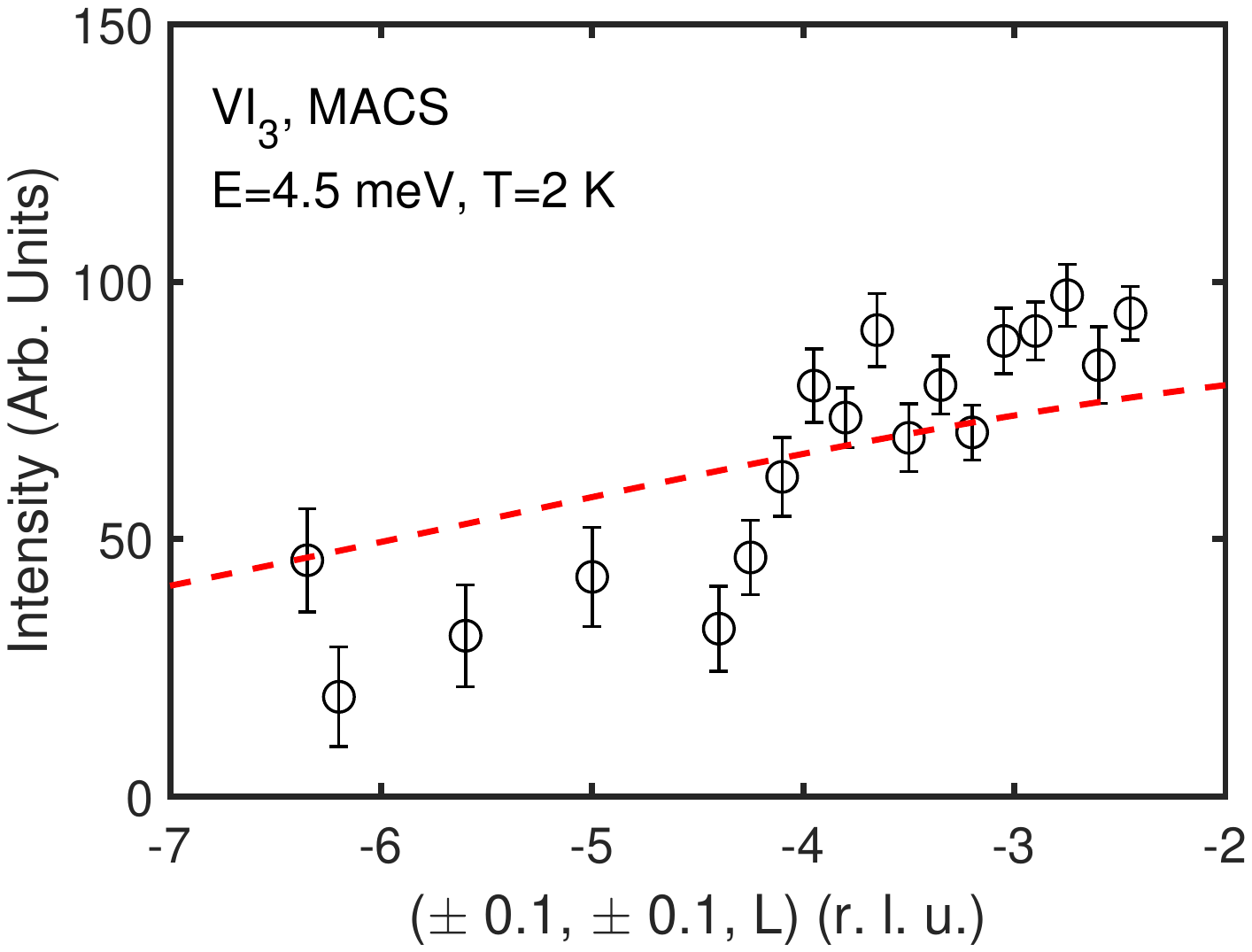}
	\caption{\label{form_factor} The $L$ dependence of the background corrected scattering in VI$_{3}$ at E=4.5 meV.  The free-ion magneitc form factor is plotted by the dashed red line.}
\end{figure}

$\textit{MACS:}$  On MACS, we utilized the broad detector coverage to measure a background and then subtract it from the data.  The underlying assumption is that dominant background is independent of the sample rotation angle (denoted as $A3$) and is outlined in Fig. \ref{MACS_back}.  While this underlying assumption may seem to be suprising given the complex nature of the plate-like sample geometry, this assumption was experimentally valid for the region around the (0,0,L) points in reciprocal space and is shown in Fig. \ref{MACS_back} $(a)$ which plots the raw data for a constant energy scan at an energy transfer of 4.5 meV.  The scan illustrates a rod of magnetic intensity along (0,0,L) and a background which is dependent on $|\vec{Q}|$, but is independent at a fixed $|\vec{Q}|$.  Utilizing this, we then removed a region around (0,0,L) and kept the background region in Fig. \ref{MACS_back} $(b)$ to form a background for all momentum transfers by assuming the background only depended on $A4$, or the scattering angle \cite{Pasztorova}.  The averaged generated background is shown in Fig \ref{MACS_back} $(c)$ and the subtracted data, used in the main text is shown in Fig. \ref{MACS_back} $(d)$.  This routine was applied to all energies measured with the constant momentum scans in the main text being a compilation of a series of constant energy scans like that shown in Fig. \ref{MACS_back}. \\
 
 We note that this methodology differs from the usual process on MACS of removing the sample and then measuring the background on an empty can or empty cryostat.  We found this method did not work and gave erratic results in this case presumably given the complex nature of the scattering of the sample+Fomblin+plate geometry.  This was likely made worse by the incoherent scattering present from both the Vanadium and also the Fomblin grease.\\

$\textit{MAPS:}$  On MAPS, owing to kinematics, we had to rotate the sample orientation, despite its two dimensional nature. Normally, when MAPS is used to measure single crystals of low-dimensional magnets only a single sample orientation is needed \cite{Horace}. However, this results in a coupling between the component of momentum along $k_{i}$ (usually chosen to be the axis perpendicular to the 2-d plane) and energy transfer. Rotating the sample decouples these two variables and allows access to smaller values of $|L|$, which can be advantageous if the intensity of the signal decays quickly due to the form factor. Given the number of aluminum plates required to obtain the necessary sample mass, and a scan range that was asymmetric around the nominal zero angle (when (0,0,L) is parallel to $k_{i}$), this introduced a large asymmetry in the background. To obtain an estimate of the background, we fit constant energy scans of the MAPS data to two Gaussians symmetrically displaced from the $\vec{Q}$=0 position plus the following background. \\

\begin{subequations}
\begin{gather}
I_{background}=
\begin{cases}
      A+B_{neg}|H|^{2}, & \text{if}\ (H,H)<0 \\
      A+B_{pos}|H|^{2}, & \text{if}\ (H,H) >0
    \end{cases} \nonumber
\end{gather}
\end{subequations}

\begin{figure}[h]
\includegraphics[trim=1cm 2cm 1cm 2cm,width=0.6\textwidth]{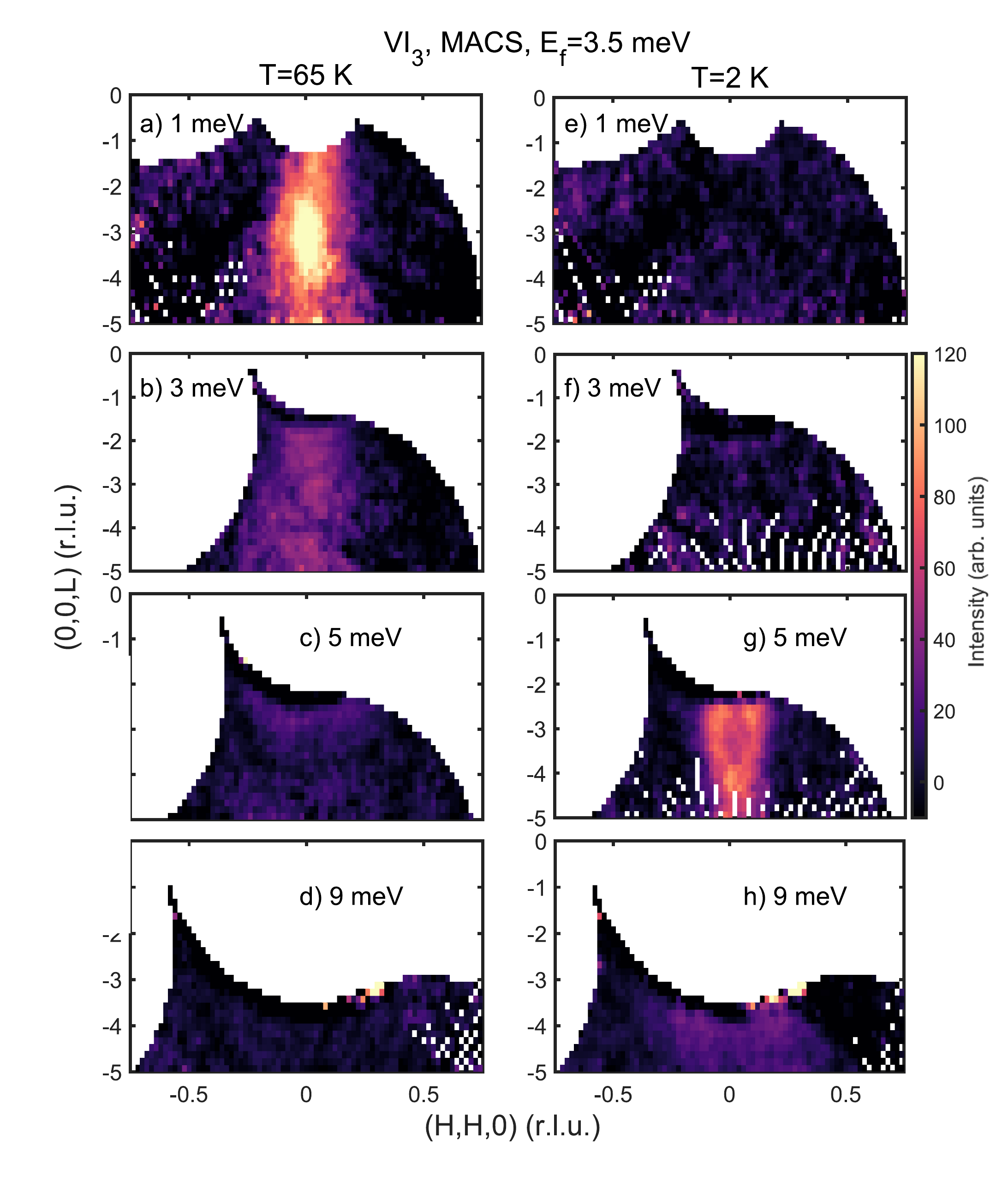}
\caption{\label{temp_constE} A comparison of a series of constant energy scans at 2 and 65 K.  Illustrative constant energy scans at 65 K are shown in $(a-d)$ and compared against the low temperature T=2 K data in $(e-h)$.}
\end{figure}

\noindent Note that this background corresponds to a parabola which is continuous at $\vec{Q}$=0, but with different concavities for negative and positive values of (H,H) to model the different geometries for a scattering beam with $2\theta$ having different sign.\\

The underlying assumption of this is that background scales as a constant plus a component that increases as $Q^{2}$ as expected for scattering from an overall instrument background and phonons from the sample.  We note that we did not consider a difference between the background along the vertical direction.  The results of this subtraction are illustrated in Fig. \ref{Back_MAPS} which plots the raw data in Fig. \ref{Back_MAPS} $(a)$ and the generated background from this fitting procedure in $(b)$.  The resulting subtracted data is shown in $(c)$.   

\section{Supplementary spectroscopic data}

In this section, we provide some additional spectroscopic data to support the data presented in the main text.  To confirm the magnetic nature of the scattering, we performed two tests.  First, we confirmed the scattering decayed in intensity with increasing momentum transfer and this is illustrated in Figure 2 of the main paper.  This is shown in Fig. \ref{form_factor} for data taken at E=4.5 meV on MACS with the free-ion V$^{3+}$ isotropic form factor plotted for comparison.  Second, we performed additional measurements on MACS (NIST) at T=65 K.  The constant energy scans from this experiment are shown in Fig. \ref{temp_constE} and illustrate that at high temperatures the magnetic scattering fills in at low energies, as expected for magnetic scattering and also for the response above the magnetic ordering temperature.\\

\begin{figure}[h]
\includegraphics[trim=0cm 1cm 0cm 0cm,width=0.6\textwidth]{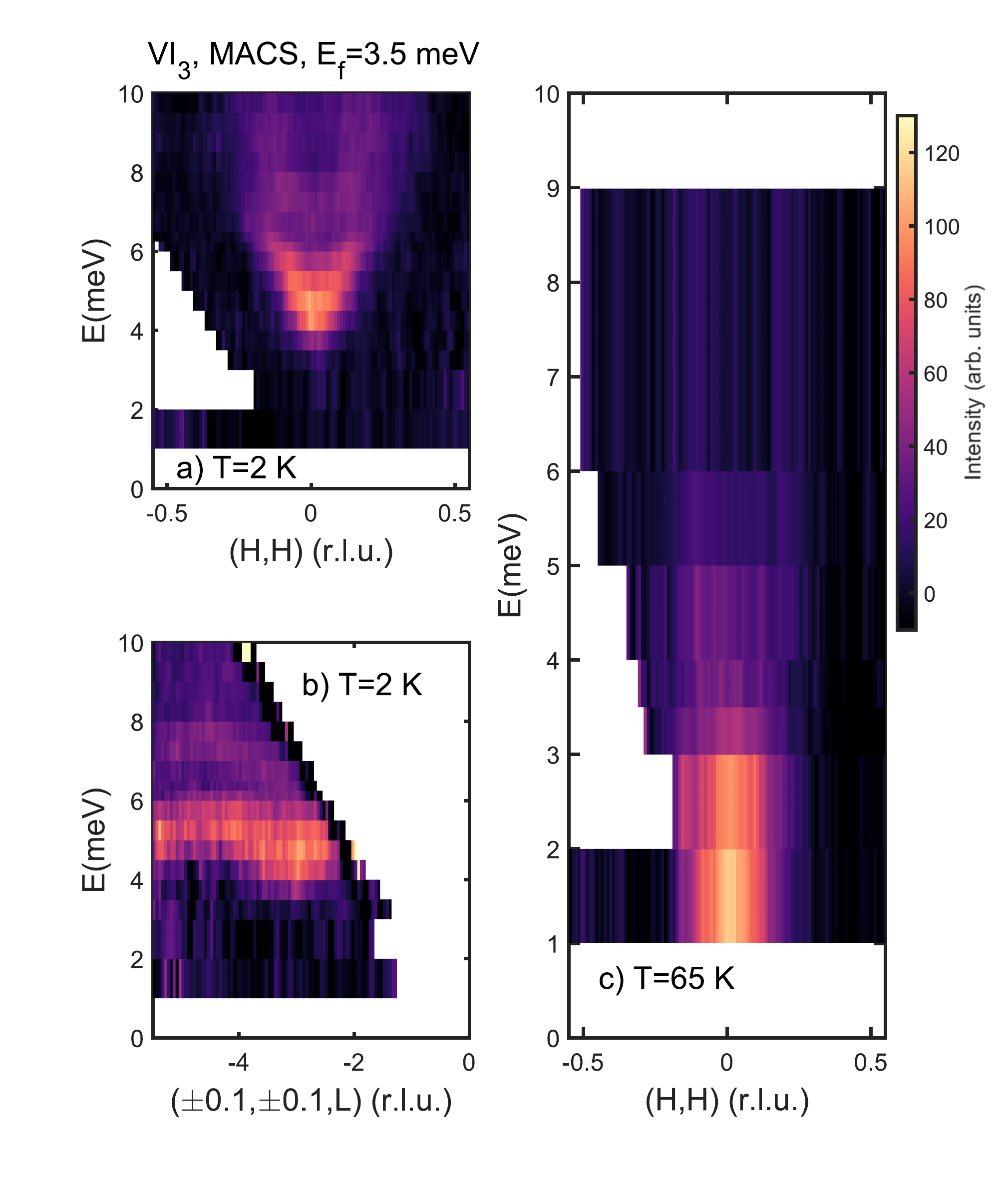}
\caption{\label{MACS_SI} Supplementary neutron inelastic scattering data is shown to support the data presented in the main text.  $(a)$ reproduces the constant momentum slice at low temperatures shown in the main text.  $(b)$ shows a constant momentum slices along the $L$ direction, suggestive of a very weak coupling of the magnetic VI$_{3}$ layers.  Panel $(c)$ shows the same constant momentum slice but at 65 K showing that the gapped ferromagnetic excitations in $(a)$ replaced by a gapless paramagnetic response.}
\end{figure}

The high temperature data is further compared against the low-temperature response in Fig. \ref{MACS_SI}.  Panel $(a)$ shows a the $L$-integrated constant momentum cut presented in the main text of the paper and is compared to the same constant momentum cut in panel $(c)$.  Owing to kinematics and the fact that the accessible (0,0,L) values change with energy transfer, the integrated range in $L$ was performed in scattering angle $A4$ and $L$ values between $A4$=18$^{\circ}$ and 40$^{\circ}$ were integrated.  At T=65 K, the gapped excitations present at low temperatures in the ferromagnetically ordered phase are replaced by a paramagnetic response as expected in the absence of a molecular field in the excitonic model discussed both in the main paper and also further outlined below. The temperature dependence is different from expectations of phonon scattering and, combined with the momentum dependence indicative of an underlying magnetic form factor, is consistent with a magnetic origin.\\

\begin{figure}[h]
\includegraphics[width=0.9\textwidth]{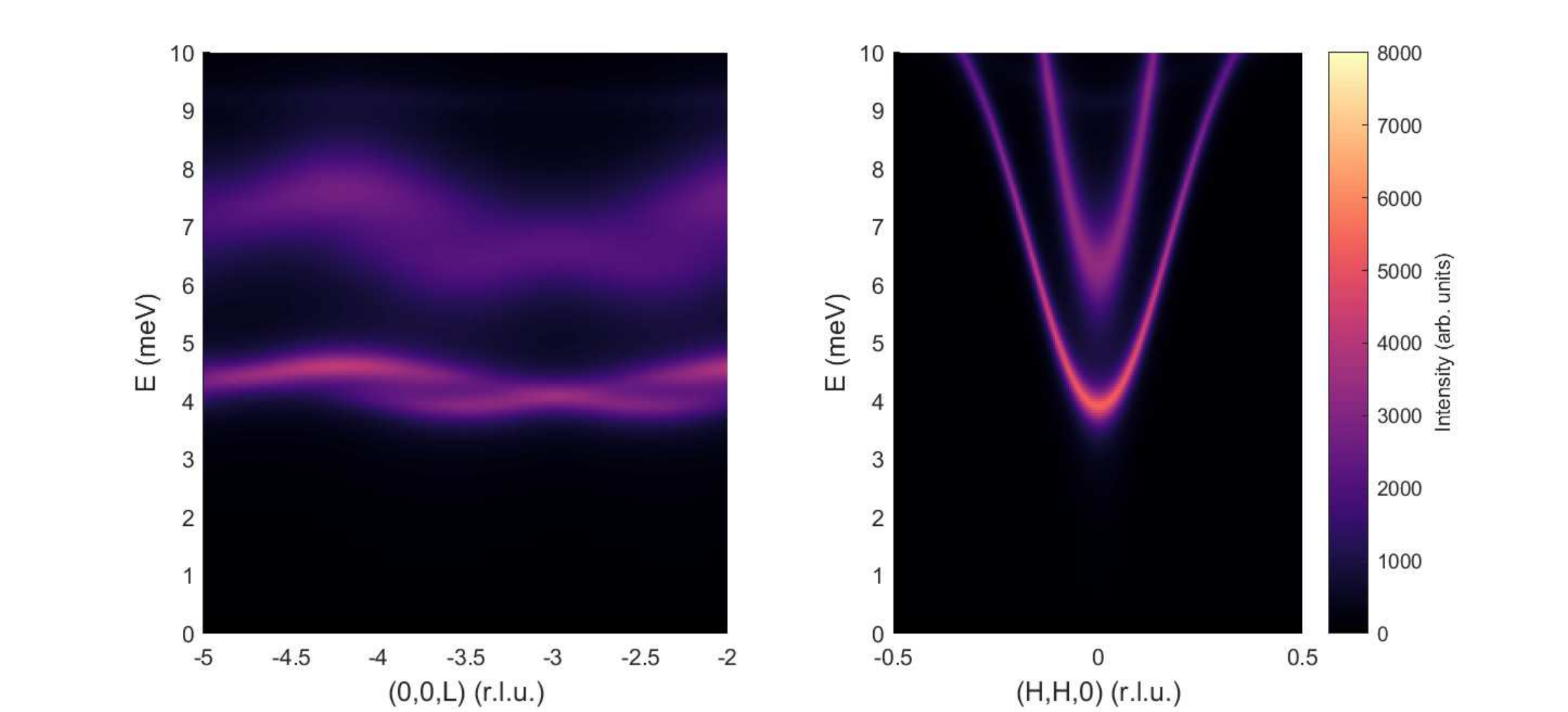}
\caption{\label{Jc} $(a)$ Calculated dispersion along L with the fitted values from the main text and a small coupling along $c$ ($J_{c}$= -0.5 meV). $(b)$ The calculated in-plane dispersion with coupling along $c$, showing a small change in the gap to the lower mode ($\sim 0.4$ meV), but no significant change in the spectrum. }
\end{figure}

A further point illustrated in Fig. \ref{MACS_SI} $(b)$ is the momentum dependence of the low temperature scattering along the (0, 0, L) direction sensitive to the coupling along $c$.  For a purely two dimensional material, no dependence would be observable.  It is interesting in Fig. \ref{MACS_SI}  that the data are consistent with this, but shows a small $\sim$ 0.5 meV dispersion along L indicative of a very weak coupling along $c$.  In terms of exchange constant, this would correspond to a coupling along $c$ of $\sim$0.5 meV.  This value is nearly within the resolution of the measurement and further confirms the two dimensional nature of the magnetic coupling and also dimensionality of the material. However, it is interesting to note that the data are suggestive of a very weak, but still finite, coupling along $c$ despite only weak van der Waals forces connecting the hexagonal VI$_{3}$ sheets. The effect of the inclusion of a small out-of-plane coupling on the in-plane response is small (Fig. \ref{Jc})) and hence is omitted from the analysis in the main text. 
\begin{figure}[h]
\includegraphics[width=0.75\textwidth]{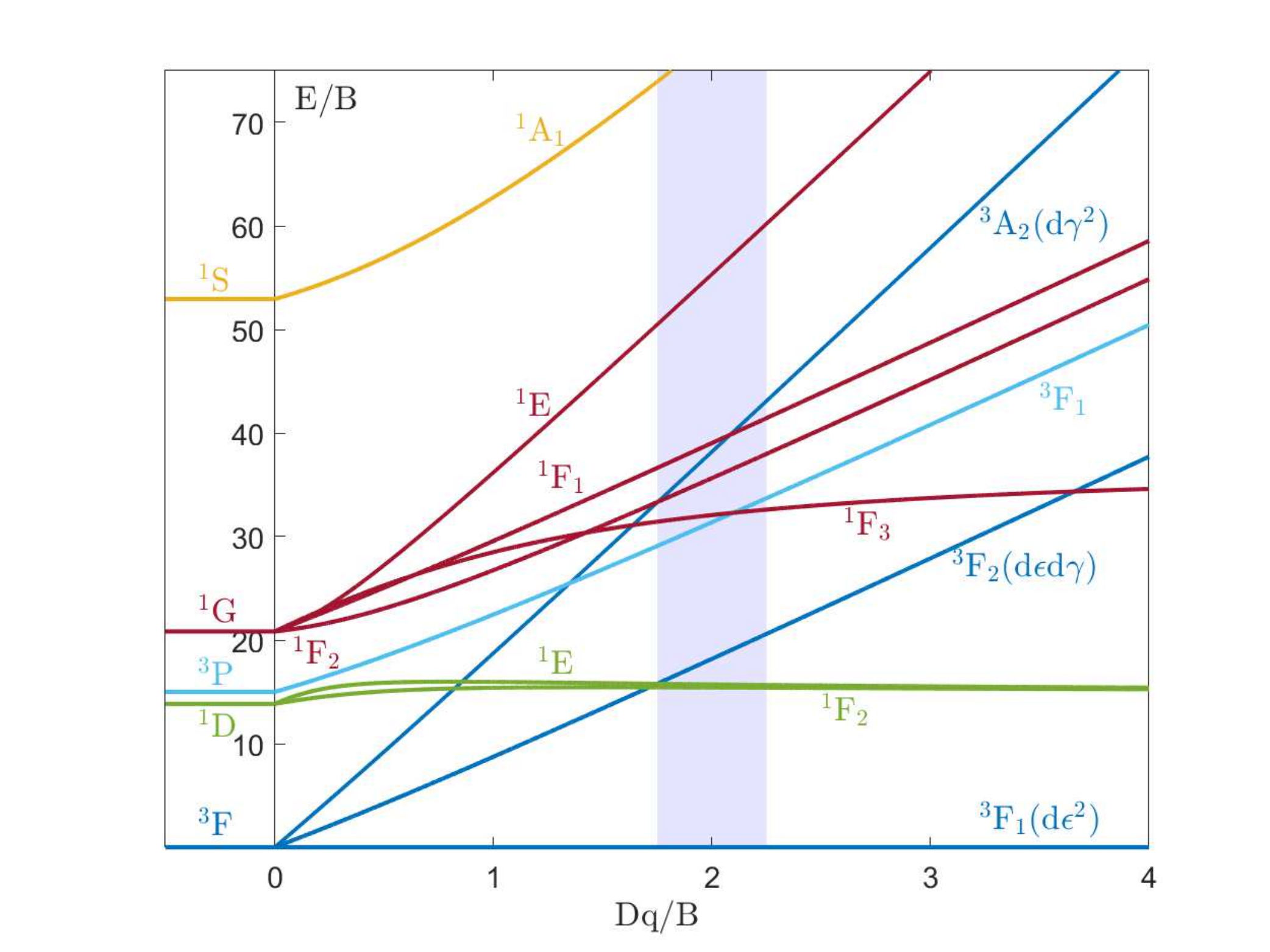}
\caption{\label{TS} Tanabe-Sugano diagram for a $d^{2}$ ion showing the splitting due to an octahedral field, with a $C/B=4.42$ as appropriate for V$^{3+}$. Corresponding strong crystal field bases are labeled in parentheses. The blue shaded region indicates the ratio Dq/B$\approx2$ that one expects for V$^{3+}$ \cite{McClure}. }
\end{figure}
\section{Intermediate Crystal Field Approach}

Previous studies \cite{Yang_Khomskii,Huang} have approached VI$_{3}$ using the strong crystal field approach, whereby the crystal field splitting is treated first and splits the five-fold orbital degeneracy into a ground state triplet, $t_{2g}$, and excited doublet, $e_{g}$. The system is written in terms of real orbital basis states \cite{Khomskii} and the ground state triplet can be projected onto an effective $l=1$ manifold, with an appropriate projection factor \cite{Tchernyshyov}, $\alpha=-1$, as with the intermediate field approach. In either picture, by performing this projection, one neglects the mixing of higher excited levels, either from the $e_{g}$ levels in the strong crystal field approach or the $^{3}P$ level in the case of the intermediate crystal field. The ground state triplet can be described in each limit as \cite{AbragamBleaney}
\begin{equation}
\mathrm{cos}\alpha \ket{t_{2}^{2}}-\mathrm{sin}\alpha \ket{t_{2}e} = \epsilon \ket{^{3}F}+\tau \ket{^{3}P}
\end{equation}  

\noindent where the $\ket{t_{2}e}$ state results from the promotion of one electron to the $e_{g}$ level. The phase, $\alpha$, can be written as $\alpha=\mathrm{arctan}[12B/(9B+10Dq)]/2$ and hence $\tau=(\mathrm{cos}\alpha - 2\mathrm{sin}\alpha)/\sqrt{5}$. One can calculate the ratios of the contribution to the ground state triplet from the lower and excited orbital levels for the strong $(0.98:-0.19)$ and intermediate $(0.96:0.27)$ crystal field approaches. In both cases, the dominant contribution is from the lower orbital level, but there exists a non-vanishing weighting from the excited state. Omission of this this mixing results in a miscalculation of the projected spin orbit coupling constant $\lambda$, though the error is of the order $\tau^{2} \approx 0.07$ \cite{AbragamBleaney}. The similarity of the admixture amplitude of the excited orbital level in both the intermediate and strong crystal field pictures suggests that V$^{3+}$ can be seen as existing on the boundary of the regions of validity of the strong and intermediate crystal field pictures, with both approaches being justifiable depending on the nature of calculation undertaken.    
\section{Exciton Model}
In the intermediate crystal field approach, the Coulomb splitting is treated first and one considers only transitions within the ground state multiplet. In the case of a $d^{2}$ ion, this is the $^{3}F$ level (Fig. \ref{TS}) which is separated in energy from the excited $^{3}P$ ion by $10Dq$ \cite{AbragamBleaney}.

The $^{3}F$ level is further split by an octahedral field
\begin{equation}
\mathcal{H}_{CF}=B_{4}\left(\mathcal{O}_{4}^{0}+5\mathcal{O}_{4}^{4}\right),
\label{CEF}
\end{equation} 

\noindent into a ground state triplet, $\Gamma_{4}$, an excited triplet $\Gamma_{5}$ and a singlet $\Gamma_{2}$,

\begin{equation}
\mathcal{H}_{CF}=B_{4}
\begin{pmatrix}
360 & 0 & 0 & 0 & 0 & 0 & 0 \\
0 & 360 & 0 & 0 & 0 & 0 & 0 \\
0 & 0 & 360 & 0 & 0 & 0 & 0 \\
0 & 0 & 0 & -120 & 0 & 0 & 0 \\
0 & 0 & 0 & 0 & -120 & 0 & 0 \\
0 & 0 & 0 & 0 & 0 & -120 & 0 \\
0 & 0 & 0 & 0 & 0 & 0 & -720 
\end{pmatrix}
\end{equation}

\noindent where $B_{4}<0$. If the crystal field splitting is large compared to the energy scales of the single-ion Hamiltonian, but small compared to the Coulomb splitting, the angular momentum operator can be projected onto the ground state triplet, $\Gamma_{4}$, by way of a projection factor, $\mathbf{L}=\alpha \mathbf{l}$ \cite{AbragamBleaney}. This reduces the size of Hamiltonian from $21\times 21$ $(S=1, L=3)$ to $9\times 9$ $(S=1, l=1)$. The projection factor can be calculated by projecting the orbital angular momentum operator into the ground state of the crystal field Hamiltonian (Eq. \ref{CEF}),

\begin{equation}
\mathcal{C}_{CF}^{-1}\hat{L}_{z}\mathcal{C}_{CF}=
\begin{pmatrix}
   -1.5&0&0& -1.9365&0&0&0\\ 
   0&0&0& 0&0&0&0\\
   0&0&1.5& 0&0&1.9365&0\\ 
   -1.9365&0&0&-0.5&0&0&0\\
   0&0&0&0&0&0&2\\
   0&0&1.9365&0&0&0.5&0\\
   0&0&0&0&2&0&0\\
\end{pmatrix}
\end{equation}
\noindent where $\mathcal{C}_{CF}$ is the matrix whose columns are the eigenvectors of the crystal field Hamiltonian. By reading off the prefactor of the Pauli matrix in the upper left block, one finds that $\alpha=-\frac{3}{2}$. Using this projection factor, the single-ion Hamiltonian can be written in terms of the fictitious orbital angular momentum operator, $l=1$, 
\begin{equation}
\mathcal{H}_{SI}=\mathcal{H}_{SO}+\mathcal{H}_{dis}+\mathcal{H}_{MF}=\alpha\lambda\mathbf{l}\cdot \mathbf{S}+\Gamma \left(\hat{l}_{z}^{2}-\frac{2}{3}\right)+\sum\limits_{j}\mathcal{J}_{ij}\hat{S}_{i}^{z}\langle \hat{S}_{j}^{z}\rangle.
\end{equation}
\noindent 

The spin-orbit coupling (SOC) constant, $\lambda$, is positive for a $d^{2}$ ion. The distortion constant, $\Gamma$, is positive for an elongation and negative for a compression. The distortion term is proportional to the Steven's operator $\mathcal{O}_{2}^{0}$ and is appropriate for a tetragonal distortion along $\hat{z}$. An additional small term $\sim  \mathcal{O}_{4}^{0}$ has been neglected since its effect on the ground state $l=1$ manifold is to bring the singlet and doublet levels closer together by an amount $225B_{4}^{0}$ \cite{AbragamBleaney} but it does not break the degeneracy of the doublet. In the case of a trigonal distortion, the reference axis can be taken to be along $[111]$ leading to an undistorted crystal field Hamiltonian, $\mathcal{H}_{CEF}=-\frac{2}{3}B_{4}(\mathcal{O}_{4}^{0}+20\sqrt{2}\mathcal{O}_{4}^{3})$, with the same energy splitting. The trigonal distortion is then proportional to $\mathcal{O}_{2}^{0}$ as in the tetragonal case, where once again we neglect the small energy renormalization from the fourth order term. The final term, $\mathcal{H}_{MF}$, represents the mean molecular field felt by the each ion due to the coupling to surrounding ions. First, treating the SOC, that the ground state is split into three levels according to the effective total angular momentum, $\mathbf{j}=\mathbf{l}+\mathbf{S}$. By diagonalizing $\mathcal{H}_{SO}$ and writing $\hat{j}_{z}$ in terms of the eigenstates of the spin-orbital Hamiltonian,

\begin{equation}
\mathcal{C}_{SO}^{-1}\hat{j}_{z}\mathcal{C}_{SO}=
\begin{pmatrix}
   2&0&0&0&0&0&0&0&0\\ 
   0&1&0&0&0&0&0&0&0\\ 
   0&0&0&0&0&0&0&0&0\\ 
   0&0&0&-1&0&0&0&0&0\\ 
   0&0&0&0&-2&0&0&0&0\\ 
   0&0&0&0&0&1&0&0&0\\ 
   0&0&0&0&0&0&0&0&0\\ 
   0&0&0&0&0&0&0&-1&0\\ 
   0&0&0&0&0&0&0&0&0\\ 
\end{pmatrix}
\end{equation}
\noindent one finds that the three levels correspond to $j_{eff}=2,1,0$ states. 

The full Hamiltonian, with inter-ion Heisenberg exchange can be written as
\begin{subequations}
\begin{gather}
\mathcal{H}=\mathcal{H}_{1}+\mathcal{H}_{2}\\
\mathcal{H}_{1}=\sum\limits_{i}\mathcal{H}_{SO}(i)+\sum\limits_{i}\mathcal{H}_{dis}(i)+\sum\limits_{i}\hat{S}_{i}^{z}\left(\sum\limits_{j}\mathcal{J}_{ij}\langle \hat{S}_{j}^{z}\rangle\right)\\
\mathcal{H}_{2}=\sum\limits_{ij}\frac{\mathcal{J}_{ij}}{2}\left(\hat{S}_{i}^{+}\hat{S}_{j}^{-}+\hat{S}_{i}^{-}\hat{S}_{j}^{+}\right)+\sum\limits_{i>j}\mathcal{J}_{ij}\hat{S}_{i}^{z}\left(\hat{S}_{j}^{z}-2\langle \hat{S}^{z}_{j}\rangle\right).
\end{gather}
\end{subequations}
\noindent One can diagonalize the single-ion part first, $\mathcal{H}_{1}=\sum\limits_{im}\omega_{m}\hat{c}_{m}^{\dagger}(i)\hat{c}_{m}(i)$, and write the spin operators in terms of the eigenvectors of the single-ion Hamiltonian, $S_{\alpha mn}=\bra{m}\hat{S}^{\alpha}\ket{n}$ \cite{Buyers}.

The equation of motion for the Green's function, $G^{\alpha\beta}_{ij}(t)=G_{ij}(\hat{S}^{\alpha},\hat{S}^{\beta},t)=-i\Theta(t)\langle [\hat{S}^{\alpha}_{i}(t),\hat{S}^{\beta}_{j}]\rangle$, can be written as 
\begin{equation}
i\partial_{t}G_{ij}(\hat{S}^{\alpha},\hat{S}^{\beta},\omega)=\delta(t)\langle [\hat{S}^{\alpha}_{i}(t),\hat{S}^{\beta}_{j}]\rangle\\ -i\Theta(t)\langle [i\partial_{t}\hat{S}^{\alpha}_{i}(t),\hat{S}^{\beta}_{j}]\rangle.
\end{equation}
\noindent Utilizing the Heisenberg equation of motion, the time-dependent spin operator in the second term can be replaced with a commutator, and after a temporal Fourier transform, one finds
\begin{equation}
\omega G_{ij}(\hat{S}^{\alpha},\hat{S}^{\beta},\omega)=\langle [\hat{S}^{\alpha}_{i},\hat{S}^{\beta}_{j}]\rangle+G_{ij}([\hat{S}^{\alpha},\mathcal{H}],\hat{S}^{\beta},\omega).
\label{eqofmot}
\end{equation}

\begin{figure}[h]
\includegraphics[width=\textwidth]{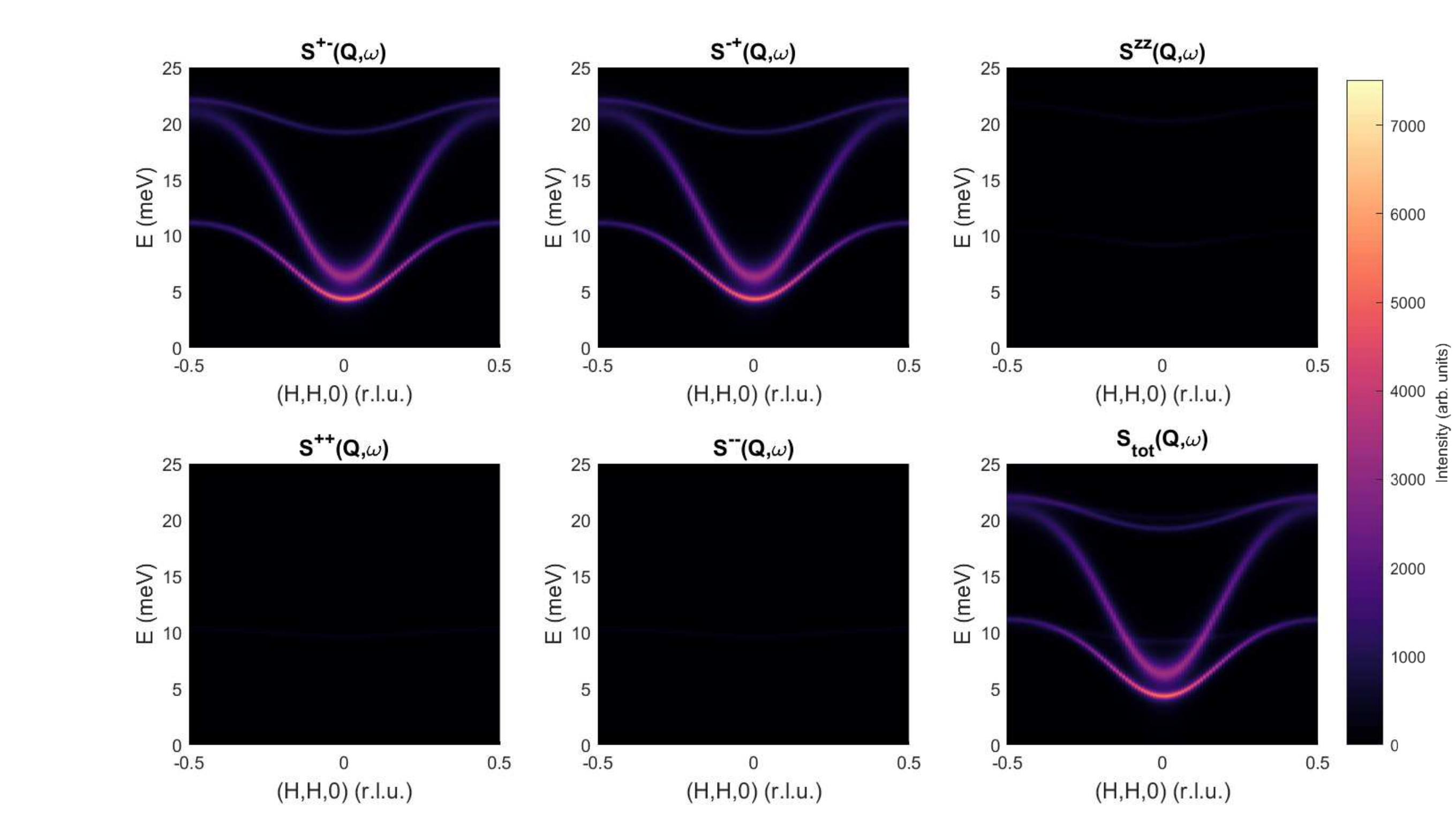}
\caption{\label{Sqw} Calculated components of the dynamic structure factor using the fitted values quoted in the main text. Negligible intensity is found for $++$, $--$ and $zz$ modes.   }
\end{figure}

To proceed, one must calculate the commutator $[\hat{S}^{\alpha},\mathcal{H}]$. This calculation is most straightforwardly done by projecting the spin operator into the ground-state multiplet of the single-ion Hamiltonian. After this transformation followed by a spatial Fourier transform, the longitudinal and transverse Green's functions can be found,
\begin{subequations}
\begin{align}
G^{+-}(\mathbf{Q},\omega)=\frac{g^{+-}(\omega)}{1-g^{+-}(\omega)\mathcal{J}(\mathbf{Q})}\\
G^{zz}(\mathbf{Q},\omega)=\frac{g^{zz}(\omega)}{1-2g^{zz}(\omega)\mathcal{J}(\mathbf{Q})},
\end{align} 
\end{subequations}
\noindent where $g^{\alpha\beta}(\omega)$ is the single-ion susceptibility 
\begin{equation}
\label{g}
g^{\alpha\beta}=\sum_{mn}\frac{\bra{m}\hat{S}^{\alpha}\ket{n}\bra{m}\hat{S}^{\beta}\ket{n}}{\omega-(\omega_{m}-\omega_{n})+i\epsilon}.
\end{equation}
The periodicity of the dispersion is given by the Fourier transform of the exchange coupling $\mathcal{J}(\mathbf{Q})=\sum_{ij}\mathcal{J}_{ij}e^{i\mathbf{Q}\cdot\left(\mathbf{r}_{i}-\mathbf{r}_{j}\right)}$. A numerical offset from the pole, $\epsilon$, is added to ensure analyticity. This offset also serves to broaden the calculated peaks, giving a finite linewidth \cite{Chou}. Owing to the difference in linewidth for the two modes, the offset was set to 0.25 for the lower mode and 0.75 for the upper mode in the MACS simulation. For the MAPS simulation this was increased to 0.5 and 1.5 respectively for the MAPS simulation to take into account the coarser energy resolution.
Introducing a site index, such that the spin on site $i$, of type $\gamma$, is $\mathbf{S}_{i\gamma}$ \cite{SarteCoO}. The Green's functions for an $n$-site system can be calculated in the same manner as for the single-sublattice problem and satisfy
\begin{subequations}
\begin{align}
G^{+-}_{\mu \nu}=&g^{+-}_{\mu \nu}+\sum_{\gamma\gamma'}\mathcal{J}_{\gamma\gamma'}(\mathbf{Q})G^{+-}_{\gamma'\nu}g^{+-}_{\mu\gamma}\\
G^{zz}_{\mu \nu}=&g^{zz}_{\mu \nu}+2\sum_{\gamma\gamma'}\mathcal{J}_{\gamma\gamma'}(\mathbf{Q})G^{zz}_{\gamma'\nu}g^{zz}_{\mu\gamma}.
\end{align}
\end{subequations}
\noindent In the random phase approximation, fluctuations on different sites are taken to be uncorrelated so that $g^{+-}_{\mu\nu}=0$, for $\mu\neq \nu$. For systems consisting of more than two sites, it is more convenient to recast the Green's function as a matrix equation
\begin{subequations}
\begin{align}
\underline{\underline{G}}^{+-}=&\underline{\underline{g}}^{+-}+\underline{\underline{g}}^{+-}\underline{\underline{\mathcal{J}}}\underline{\underline{G}}^{+-}\\
\underline{\underline{G}}^{-+}=&\underline{\underline{g}}^{-+}+\underline{\underline{g}}^{-+}\underline{\underline{\mathcal{J}}}\underline{\underline{G}}^{-+}\\
\underline{\underline{G}}^{zz}=&\underline{\underline{g}}^{zz}+2\underline{\underline{g}}^{zz}\underline{\underline{\mathcal{J}}}\underline{\underline{G}}^{zz}.
\end{align} 
\end{subequations} 
The Green's function, $G^{\alpha\beta}$, can be found by summing the $n\times n$ components of the matrices $G^{\alpha\beta}_{\mu \nu}$. Finally the total Green's function is given by
\begin{equation}
G_{tot}\left(\mathbf{Q},\omega\right)=\frac{1}{2}\left[G^{+-}\left(\mathbf{Q},\omega\right)+G^{-+}\left(\mathbf{Q},\omega\right)\right]+G^{zz}\left(\mathbf{Q},\omega\right).
\end{equation}
The dynamic structure factor is proportional to the imaginary part of the Green's function by the fluctuation-dissipation theorem, $S(\mathbf{Q},\omega)\propto-\textup{Im}[G(\mathbf{Q},\omega)]$. The components of the structure factor correspond to transverse $(S^{+-}, S^{-+})$ and longitudinal $(S^{zz})$ excitations. The components $S^{++}$ and $S^{--}$ have negligible intensity as expected for a high symmetry system on a Bravais lattice (Fig. \ref{Sqw}).

The nearest neighbor exchange, $J$, and distortion parameters for the two domains, $\Gamma_{I,II}$ were determined by fitting Gaussian peaks to one-dimensional cuts through the data using \texttt{Horace} \cite{Horace}. The calculated dispersion was then fitted to the extracted experimental dispersion curve. Below we summarize all variables used within the multi-level spin wave model and state their origin (Table \ref{Parameters}).  As discussed in the main paper, we considered the case where $B_{4}\sim$ 3.8 meV~\cite{McClure,AbragamBleaney} resulting in an orbital excited state energy of 480$B_{4}\sim$1.8 eV.  This parameter is not fitted in our analysis and its large value simply fixes the the magnetic ground state of V$^{3+}$ to be $|l=1,S=1\rangle$.

\begin{table}[h]
\caption{\label{Parameters} Parameters used in the multi-level spin wave model.}
\begin{ruledtabular}
\begin{tabular}{cccc}
Parameter &Value (meV) &Origin \\ 
\hline
J & -8.6 ($\pm$0.3)    & Fitted from experiment \\ 
$\Gamma_{I}$ & 3.4 ($\pm$0.02) & Fitted from experiment \\ 
$\Gamma_{II}$ & -13.7 ($\pm$0.5) & Fitted from experiment \\ 
$\lambda$ & 12.9 & Literature \citep{AbragamBleaney} \\ 
\end{tabular}
\end{ruledtabular}
\end{table}   


%